\documentclass[twocolumn,prl,floatfix,citeautoscript,nofootinbib,superscriptaddress]{revtex4}
\usepackage{amsbsy}
\usepackage{latexsym,epsfig,graphicx}
\usepackage{dcolumn}
\usepackage{graphicx}
\usepackage{subfigure}
\usepackage{comment}
\usepackage{color}
\usepackage{bm}
\usepackage{mathrsfs}
\usepackage{amsfonts}
\usepackage{amsmath}
\usepackage{color}
\usepackage{amssymb}
\usepackage{xspace}
\usepackage{epstopdf}
\usepackage{tabularx}
\usepackage{longtable}
\usepackage[colorlinks=true, letterpaper=true, pdfstartview=FitV, linkcolor=blue, citecolor=blue, urlcolor=blue]{hyperref}
\usepackage[normalem]{ulem}

\pdfoutput=1

\begin{document}

\title{Topological bands and triply-degenerate points in non-Hermitian
hyperbolic metamaterials}
\author{Junpeng Hou}
\affiliation{Department of Physics, The University of Texas at Dallas, Richardson, Texas
75080-3021, USA}
\author{Zhitong Li}
\affiliation{Department of Electrical and Computer Engineering, The University of Texas
at Dallas, Richardson, Texas 75080-3021, USA}
\author{Xi-Wang Luo}
\email{xiwang.luo@utdallas.edu}
\affiliation{Department of Physics, The University of Texas at Dallas, Richardson, Texas
75080-3021, USA}
\author{Qing Gu}
\affiliation{Department of Electrical and Computer Engineering, The University of Texas
at Dallas, Richardson, Texas 75080-3021, USA}
\author{Chuanwei Zhang}
\email{chuanwei.zhang@utdallas.edu}
\affiliation{Department of Physics, The University of Texas at Dallas, Richardson, Texas
75080-3021, USA}

\begin{abstract}
Hyperbolic metamaterials (HMMs), an unusual class of electromagnetic
metamaterials, have found important applications in various fields due to
their distinctive properties. A surprising feature of HMMs is that even
continuous HMMs can possess topological edge modes. However, previous
studies based on equal-frequency surface (analogy of Fermi surface) may not
correctly capture the topology of entire bands. Here we develop a
topological band description for continuous HMMs that can be described by a
non-Hermitian Hamiltonian formulated from Maxwell's equations. We find two
types of three dimensional non-Hermitian triply-degenerate points with
complex linear dispersions and topological charges $\pm 2$ and 0 induced by
chiral and gyromagnetic effects. Because of the photonic nature, the vacuum
band plays an important role for topological edge states and bulk-edge
correspondence in HMMs. The topological band results are numerically
confirmed by direct simulation of Maxwell's equations. Our work presents a
general non-Hermitian topological band treatment of continuous HMMs, paving
the way for exploring interesting topological phases in photonic continua
and device implementations of topological HMMs.
\end{abstract}

\maketitle


{\color{blue}\emph{Introduction}}. Hyperbolic metamaterials (HMMs), also
known as indefinite media, are a class of optical metamaterials with extreme
anisotropy \cite{SmithDR2003}: the effective permittivity (or permeability)
tensor components that are parallel and perpendicular to the optical axis
have opposite signs, therefore their optical properties resemble dielectric
and metal in orthogonal directions \cite{SmithDR2003,PoddubnyA2013}. Due to
such unique property and associated indefinite dispersion, HMMs possess
infinite optical density of states, giving rise to applications in versatile
fields \cite%
{JacobZ2006,LiuZ2007,KabashinAV2009,ShoaeiM2015,GalfskyT2016,LuD2014,FerrariL2017,ChandrasekarR2017}
such as super-resolution microscopy, biosensing, lasing, etc.

Recently, it was proposed \cite{LiuC2017,GaoW2015,ChernRL2008} that HMMs can
serve as an ideal candidate for studying topological photonics in materials
with continuous translational symmetry (\textit{i.e.}, no periodic lattice
structure at optical wavelength scale or the periodicity goes to infinity)
\cite{note}. Topological photonics, the application of topological band
theory in photonic systems, have generated great excitements for both
fundamental studies and practical applications. Most studies have focused on
periodic dielectric systems \cite{OzawaT2018} (e.g., photonic crystals,
coupled waveguides and cavities), which are well described by band topology
in Bloch basis based on the analogy between electromagnetic wave equations
and the Schr\"{o}dinger's equation \cite%
{LuL2014,KhanikaevAB2017,HaldaneFDM2008,RaghuS2008,WangZ2009,RamanA2010,
LuL2010,WangL2016,SirokiG2017,BahariB2017,GaoZ2017,PartoM2018,XiaoM2016}.

Different from Hermitian dielectric systems \cite%
{HaldaneFDM2008,RaghuS2008,SilveirinhaMG2015} with real-valued band
structures, HMMs represent a continuous non-Hermitian system with complex
eigenvalues due to their metal nature along one or two of the optical axes.
Therefore two important questions naturally arise. Can a theory be developed
for characterizing topological bands of such continuous non-Hermitian HMMs?
If so, what new physics can arise from such topological band theory? We note
that previous studies have introduced the equal frequency surface (EFS) to
characterize the topology of HMMs \cite{LiuC2017,GaoW2015,ChernRL2008}, with
photonic EFS corresponding to the Fermi surface in electronic materials.
While the Fermi surface does contain certain information, the complete
topological properties are encoded in the entire bands. As a result, the EFS
theory is incomplete for investigating the topological properties of
continuous non-Hermitian HMMs, and may lead to ambiguous (sometimes
misleading or incorrect) predictions (see Supplementary Materials (SM) \cite%
{Supp} for an example).

In this Letter, we answer these two important questions by developing a
topological band description, along with the bulk-edge correspondence, for
continuous HMMs. Our main results are:

\textit{i}) An effective non-Hermitian Hamiltonian for HMMs is derived from
Maxwell's equations. Symmetry analysis shows the physics can be described by
three bands (\textit{i.e.}, a spin-1 system). Proper gyromagnetic or chiral
field opens a band gap between the upper and the other two bands except at $%
\mathbf{k}=0$, which is a non-Hermitian triply-degenerate point (TDP) \cite%
{TDP1,TDP2,HuH2018} with complex linear band dispersions (\textit{i.e.}, a
topological semimetal). The complex bulk spectrum exhibits an exceptional
cone with the TDP as cone vertex. TDPs were studied recently in solid state
\cite{TDP1,TDP2} and ultracold atomic systems \cite{HuH2018}, but have not
been explored in photonic materials or any non-Hermitian systems, and their
real linear dispersions are very different from non-Hermitian TDPs. The
topological charge of the TDP at $\mathbf{k}=0$ is $\pm 2$ (0) for chiral
(gyromagnetic) effect. For any fixed nonzero $k_{z}$, the HMM is a 2D Chern
insulator, and the TDP emerges as the band gap closing point at $k_{z}=0$.

\textit{ii}) There exist surface states connecting the single TDP to
infinity for both cases (change $\pm2$ or charge 0), which are illustrated
through topological edge states in both 3D and 2D Chern insulators with
fixed $k_{z}$ using the bulk-edge correspondence. More importantly, the
topological edge states can only be found in the common band gap of the HMMs
and vacuum because unlike electrons in solid-state materials, photons can
propagate in the vacuum, forming vacuum band structures outside the HMMs.
The edge states are purely real and do not suffer loss as the complex bulk,
which, combining with the unique properties of HMMs, enable the design of
novel optical devices such as topological lasing.

\textit{iii}) Our theoretical predictions on topological bands and chiral
edge states of HMMs are confirmed by numerically solving the Maxwell's
equations using COMSOL simulations.

{\color{blue}\emph{Non-Hermitian Hamiltonian and topological invariant}}.
The HMMs can be described by the source-free Maxwell's equations with the
following constitutive relation
\begin{equation}
\bm{D}=\epsilon \bm{E}+i\gamma \bm{H},~\bm{B}=\mu \bm{H}-i\gamma \bm{E},
\label{relation}
\end{equation}%
based on the symmetrized Condon set \cite{LeknerJ1996}, where $\epsilon $, $%
\mu $ and $\gamma $ are 3$\times 3$ permittivity, permeability and chirality
tensors. Without gain and loss, they satisfy $\mu ^{\dagger }=\mu $ and $%
\epsilon ^{\dagger }=\epsilon $. The chirality term can be written as $%
\gamma =\text{Tr}(\gamma )I/3+N$ with $I$ the identity matrix and $N$ a
real-valued symmetric trace-free tensor. The chiral and gyromagnetic effects
for HMMs can be induced by nonzero $\gamma $ and imaginary non-diagonal
terms in $\epsilon $ or $\mu $, respectively. The Maxwell's equations can be
recast to a linear-transformation form $H|\Psi \rangle =\omega |\Psi \rangle
$, with%
\begin{equation}
H=\left( {%
\begin{array}{cc}
\epsilon & i\gamma \\
-i\gamma & \mu%
\end{array}%
}\right) ^{-1}\left( {%
\begin{array}{cc}
0 & p \\
-p & 0%
\end{array}%
}\right) ,\text{ }|\Psi \rangle =\left( {%
\begin{array}{c}
\bm{E} \\
\bm{H}%
\end{array}%
}\right) ,  \label{MainEqu}
\end{equation}%
where $p_{[mn]}=\varepsilon _{mnl}\nabla _{l}$ is an antisymmetric tensor
operator ($p^{T}=-p$) defined through the Levi-Civita symbol $\varepsilon
_{mnl}$. In the limit $\gamma \rightarrow 0$, Eq. (\ref{MainEqu}) reduces to
the Hermitian formalism in previous works \cite%
{HaldaneFDM2008,RaghuS2008,SilveirinhaMG2015} if $\epsilon $ and $\mu $ are
positive-definite. In the context of HMMs, the Hamiltonian in Eq.~(\ref%
{MainEqu}) is generally non-Hermitian and possesses complex eigenvalues,
therefore the topological classifications for Hermitian systems \cite%
{SchnyderAP2008,KitaevA2009,RyuS2010} do not apply.

The Hamiltonian has 6 bands, which appear in pairs $\left( \omega ,-\omega
\right) $ due to the symmetry $\Pi H\Pi ^{-1}=-H$, where the symmetry
operator $\Pi $ is defined as the composite of chiral symmetry $\mathcal{C}$
and the operation $\gamma \rightarrow -\gamma $. Here $\mathcal{C}=\sigma
_{z}\otimes I_{3}$ and $\sigma _{i}$ represents Pauli matrix in the $\left( %
\bm{E},\bm{H}\right) $ space. In addition, the state at $\left( \mathbf{k}%
,\omega \right) $ represents the same physical state as that at $\left( -%
\mathbf{k},-\omega \right) $ due to the symmetry $H(-p)=-H(p)$, which holds
for arbitrary $H$. When combined together, these symmetries dictate that
only three bands are independent. Here we consider three bands with $\Re
(\omega )\geq 0$ ($\Re $ takes the real part), which form an effective
spin-1 system. Note that one band is a zero-energy ($\omega =0$) flat band,
which represents the static solutions $\bm{E}=\mathbf{\nabla }d(\bm{r})$ and
$\bm{H}=\mathbf{\nabla }b(\bm{r})$. Interestingly, the three bands are
always (triply) degenerate at $\left( \mathbf{k},\omega \right) =0$ for
arbitrary $H$, independent of $\epsilon $, $\mu $, and $\gamma $.

The energy spectra for a non-Hermitian Hamiltonian are generally complex,
and the topological invariants can be defined by either eigenvalues or
eigenstates. The eigenvalue-based winding number for a closed loop in
momentum space is defined as \cite{GongZ2018,ShenH2018} $C_{\omega }=\oint d%
\mathbf{k}\frac{\partial }{\partial _{\mathbf{k}}}\text{arg}[\omega (\mathbf{%
k})]$, which is generally trivial and irrelevant to the topological edge
modes for HMMs discussed here. On the other hand, the bands for HMMs are
separable in the complex plane, therefore winding number $W=\frac{1}{\pi }%
\oint_{\mathcal{S}^1}d\mathbf{k}\cdot \mathscr{A}(\mathbf{k})$ and the Chern
number $C=\frac{1}{2\pi}\oint_{\mathcal{S}}d\mathbf{S}\cdot \mathscr{F}$
based on eigenstates are well-defined and quantized, which can be used to
characterize the topological properties of HMMs. Here $\mathcal{S}^1$ is a
closed 1D loop and $\mathcal{S}$ can be a closed 2D sphere $\mathcal{S}^2$
(or infinite plane $\mathcal{R}^2$) in the momentum space, $\mathscr{A}(%
\mathbf{k})=-i{}_{L}\langle \Psi (\mathbf{k})|\nabla _{\mathbf{k}}|\Psi (%
\mathbf{k})\rangle _{R}$ and $\mathscr{F}=\nabla \times \mathscr{A}(\mathbf{k%
})$ are the Berry connection and Berry curvature respectively, and $|\Psi (%
\mathbf{k})\rangle _{R}$ ($|\Psi (\mathbf{k})\rangle _{L}$) is the right
(left) eigenstate \cite{ShenH2018} of the Hamiltonian. Among the three
bands, the zero-energy flat band is topologically trivial, while the other
two nonzero bands possess opposite topological invariants. Hereafter we only
plot the two nonzero-energy bands with $\Re (\omega )>0$ for better
visualization.

\begin{figure}[b]
\includegraphics[width=0.48\textwidth]{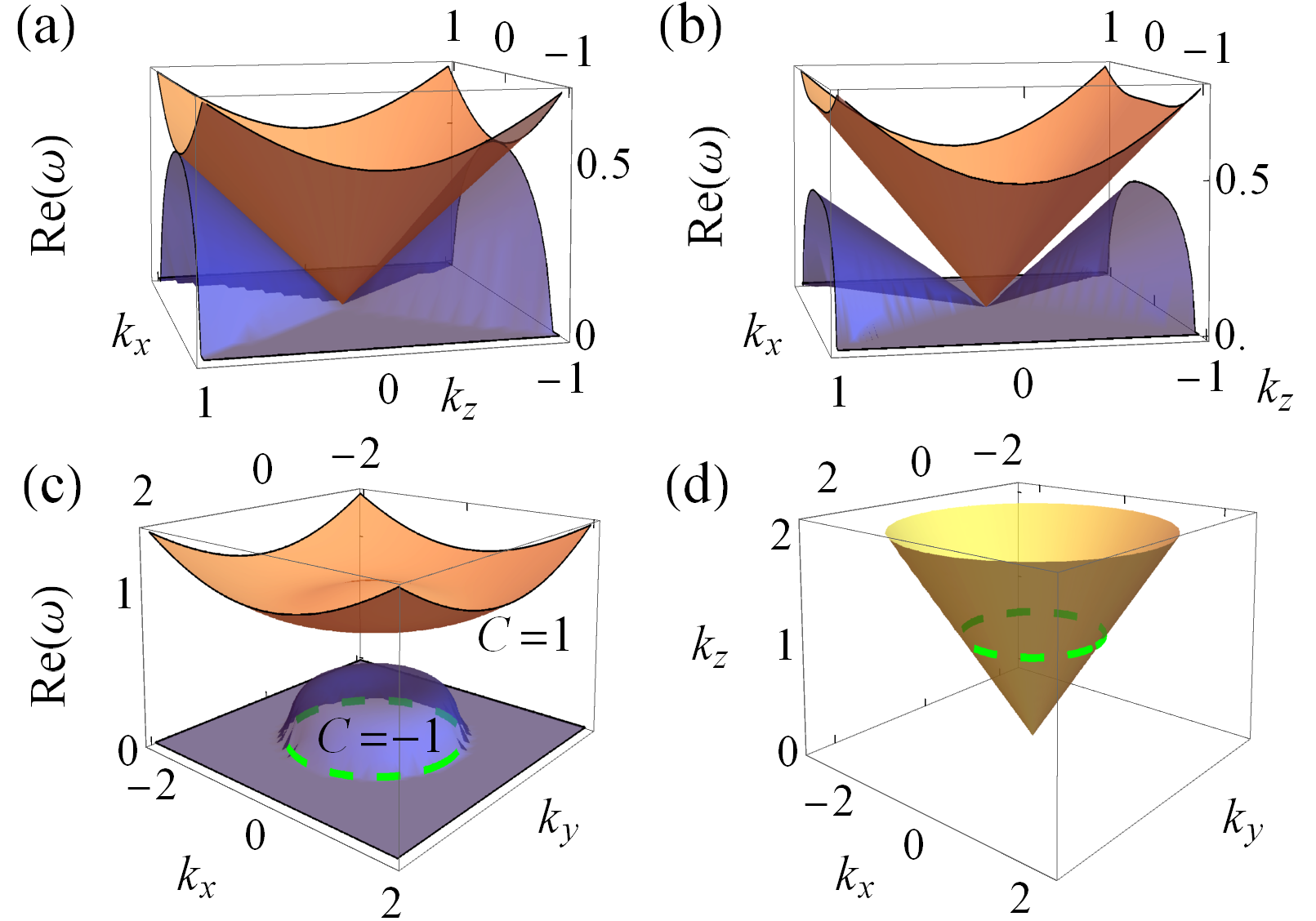} \centering
\caption{Typical band structures for HMMs. (a) A HMM with $(\protect\epsilon %
_{x},\protect\epsilon _{y},\protect\epsilon _{z})=(4,4,-3)$ exhibits a
degenerate line along $k_{z}$ axis between two nonzero bands. (b) The
degenerate line (except $\mathbf{k}=0$) in (a) is lifted by $\protect\gamma =%
\text{diag}(1,0,0)$. (c) Corresponding gapped topological bands in 2D $k_{x}$%
-$k_{y}$ plane for $k_{z}=1$. See SM \protect\cite{Supp} for the imaginary
bands. The dashed green circle is the exceptional ring. (d) The 3D
exceptional cone in momentum space at $k_z\geq0$. }
\label{fig1}
\end{figure}

{\color{blue}\emph{Charge $\pm $2\ TDPs from chiral effects}}. Without
chiral and gyromagnetic terms and assume $\epsilon =$diag$(\epsilon
_{x}>0,\epsilon _{y}>0,\epsilon _{z}<0)$ and $\mu =I$ for hyperbolic
dispersion, there is one degenerate line along the $k_{z}$ axis between the
two upper non-zero bands with $\epsilon _{x}=\epsilon _{y}$, as shown in
Figs.~\ref{fig1}(a). The degenerate line possesses a non-trivial winding
number (defined by the highest band) $W=2$ for a closed loop encircling the
line \cite{Supp}. The corresponding band structure in the $k_{x}$-$k_{y}$
plane with a fixed non-zero $k_{z}$ contains a quadratic band touching point
with winding number $W=2$ at $\left( k_{x},k_{y}\right) =\left( 0,0\right) $%
, which is computed on a closed circle enclosing the degenerate point. The
band structures for $\epsilon _{x}\neq \epsilon _{y}$ are presented in SM
\cite{Supp}.

The degeneracy between two non-zero bands along the $k_{z}$ axis (except at $%
\mathbf{k}=0$) can be lifted by breaking 
inversion symmetry using 
a chiral term (Fig.~\ref{fig1}(b) with $\gamma =$diag$(1,0,0)$). For a fixed
$k_{z}\neq 0$, the gap at the quadratic band touching point is opened,
yielding 2D Chern insulators with opposite Chern numbers $-1$ and $+1$ for $%
k_{z}<0$ and $k_{z}>0$ because the inversion symmetry along the $z$ axis is
broken (Fig.~\ref{fig1}(c)). Note here the 2D Chern number is always defined
by the upper band that is fully gapped except at $\mathbf{k}=0$. The 2D
Chern number is integrated over the 2D infinite plane $\mathcal{R}^2$ in
momentum space at constant $k_{z}$ and is quantized in continuous limit (see
SM \cite{Supp} for a proof). The lower non-zero band transits from real to
imaginary eigenenergies along an exceptional ring with coalesced eigenstates
(the green circle in Fig. \ref{fig1}(c)). Such an exceptional ring at finite
$k_{z}$ shrinks to a point at $\mathbf{k}=0$, resulting in a 3D \textit{%
exceptional cone} with the cone vertex at $\mathbf{k}=0$ (Fig. \ref{fig1}%
(d)).

\begin{figure}[t]
\centering
\includegraphics[width=0.48\textwidth]{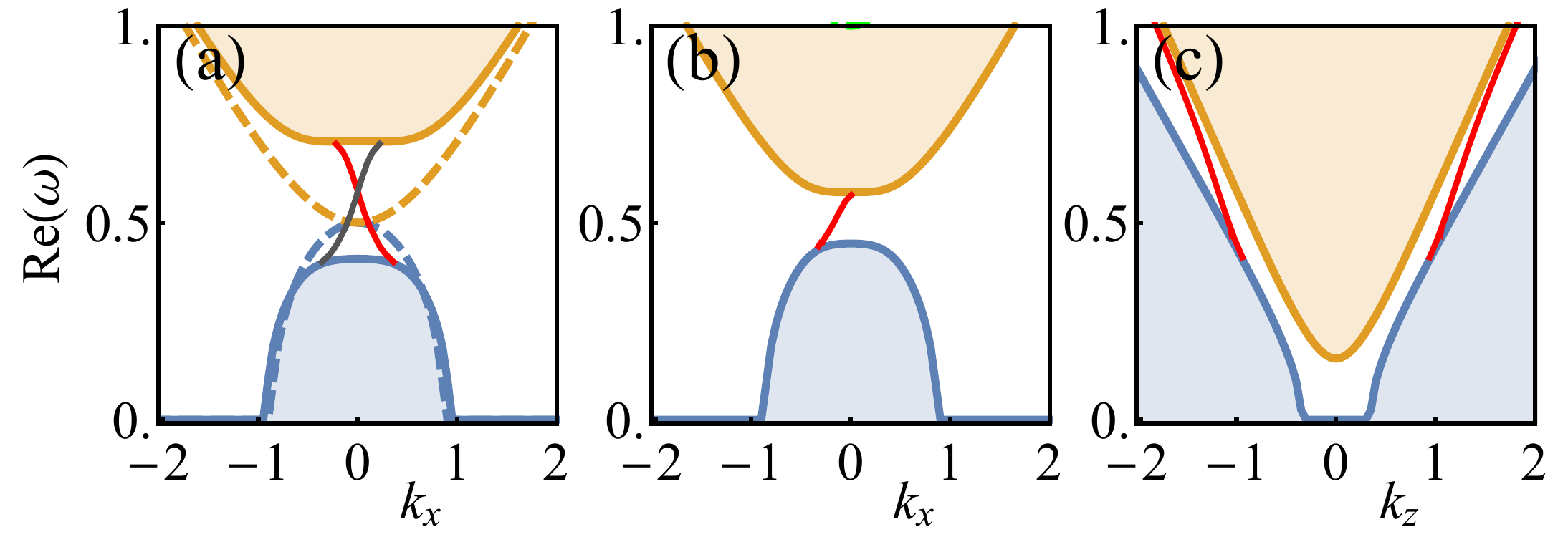}
\caption{2D band structure with edge states. We choose $(\protect\epsilon %
_{x},\protect\epsilon _{y},\protect\epsilon _{z})=(4,4,-3)$. The color-coded
areas represent sub-bands under projection and solid red (or dark gray)
curves are chiral surface waves with velocity $d\protect\omega /dk$. (a) The
edge states are induced by a pure chiral effect $\protect\gamma =\text{diag}%
(1,0,0)$ such that they possess opposite chirality for $k_{z}=+1$ (red) and $%
k_{z}=-1$ (dark gray) while the band structure remains the same. The two
dashed curves show the bands with $\protect\gamma =0$. (b) The chiral edge
states are the same at $k_{z}=\pm 1$ for a gyromagnetic term $\protect%
\epsilon _{xy}=-\protect\epsilon _{yx}=i$. (c) Same as (b) but we set $%
k_{x}=-0.3$ and compute the edge states along $k_{z}$. Here a zero charge
TDP yields two edge states with opposite chirality.}
\label{fig2}
\end{figure}

The origin $\mathbf{k}=0$ is a TDP with linear band dispersions (Fig.\ \ref%
{fig1}(b) and \cite{Supp}), which, for the lower band, can appear in either
real or imaginary spectrum along different momentum directions. Such a
non-Hermitian TDP is quite different from the real TDPs in electronic and
cold atomic Hermitian systems \cite{TDP1,TDP2,HuH2018}. At $k_{z}=0$, the
band gaps for 2D Chern insulators close, yielding a topological charge $C=+2$
of the TDP that is equivalent to the change of 2D Chern number across $%
k_{z}=0$. Here the topological charge is evaluated on a closed surface $%
\mathcal{S}^2$ enclosing $\mathbf{k}=0$. Because there is only one charge $+2
$ TDP in the HMM due to its continuous translational symmetry, there should
be surface states connecting the TDP to infinity. We consider an open
boundary condition along the $y$ direction with a semi-infinite HMM in $y<0$
and the vacuum (i.e., $\mu _{v}=\epsilon _{v}=I$) at $y>0$, and the surface
state is solved as Dyakonov wave \cite{DyakonovM1998}. Within the scope of
this work, we find that the surface wave only has real energy despite the
complex bulk spectrum. The obtained surface states in the $k_{x}$-$k_{z}$
plane connect two bulk bands and vanish at the TDP. Because the band gap
appears at different $\omega $ regions for different $k_{z}$, the commonly
used surface spectral density at a fixed $\omega $ is not good for
describing the surface states of continuous HMMs. For a fixed $k_{z}\neq 0$,
the chiral edge states propagate along opposite directions (i.e., opposite
velocities $d\omega /dk_{x}$) for $k_{z}>0$ and $k_{z}<0$ (Fig.~\ref{fig2}%
(a)) because of their opposite bulk Chern numbers of 2D insulators. Although
the lower band is purely imaginary in part of the momentum space, the edge
states only connect to purely real parts.

{\color{blue}\emph{Charge 0 TDP from gyromagnetic effects}}. The degenerate
line in Fig. \ref{fig1}(a) can also be gapped out by the gyromagnetic
effect, leading to another type of TDP at $\mathbf{k}=0$. We consider the
gyromagnetic effect that is induced by a magnetic field along the $z$
direction, which yields a pure imaginary non-diagonal term $\epsilon _{xy}$ (%
$\epsilon _{yx}=-\epsilon _{xy}$ to keep $\epsilon $ Hermitian). The
resulting band structure is similar as Fig. (\ref{fig1}b) (see Fig. (\ref%
{fig3}b)). However, the Chern numbers for 2D bands in the $k_{x}$-$k_{y}$
plane are $+1$ for both $k_{z}>0$ and $k_{z}<0$ because the magnetic field
along the $z$ direction, although breaks the time-reversal symmetry, still
preserves the inversion symmetry along the $z$ axis. The Chern number
changes sign with the sign of $\epsilon _{xy}$, i.e., sign($\Im (\epsilon
_{xy})$) ($\Im \ $takes the imaginary part). Although the band topology does
not change across $k_{z}=0$, the band gap still closes, leading a
topological TDP at $\mathbf{k}=0$ with charge 0 due to opposite Berry flux
for $k_{z}>\left( <\right) $ $0$ \cite{Supp}.

\begin{figure}[t]
\centering
\includegraphics[width=0.48\textwidth]{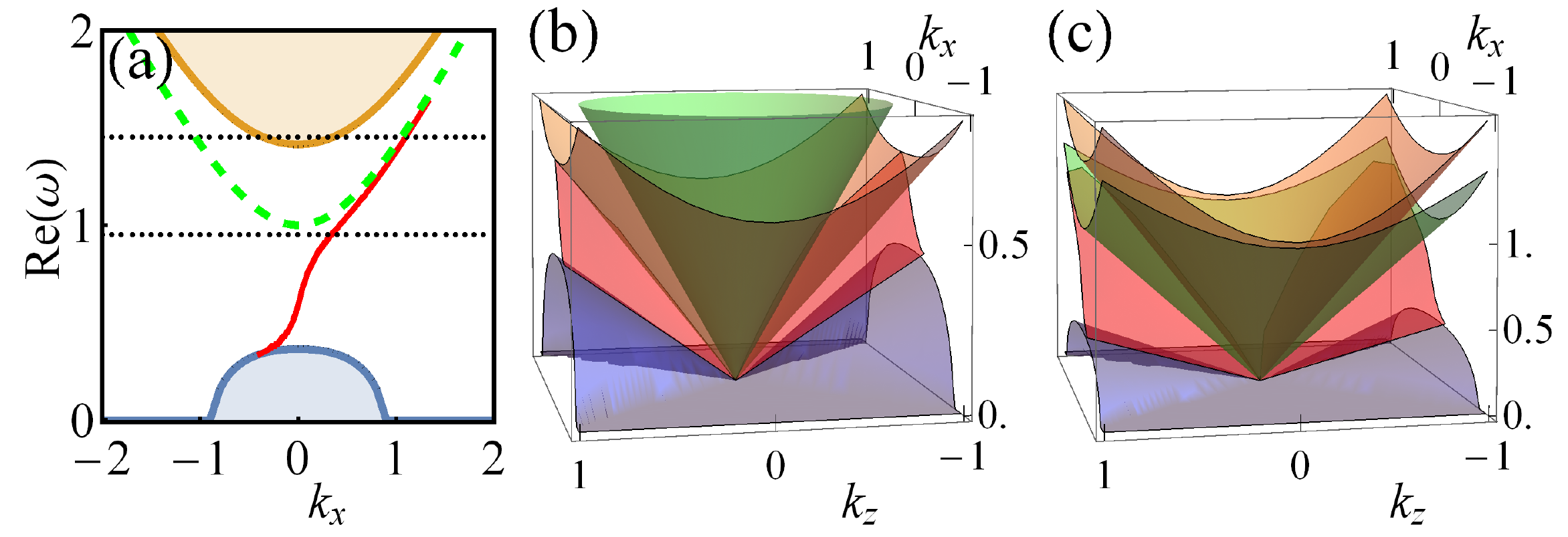}
\caption{(a) 2D band structure and edge states with a gyromagnetic term $%
\protect\epsilon _{xy}=3.5i$ for $k_{z}=1$. The dashed green curve is the
vacuum band, which is two-fold degenerate. Two dashed curves (from top to
bottom) give the frequencies of the line source in COMSOL simulations shown
in Fig.~\protect\ref{fig4}(c,a). (b,c) 3D band structures with edge states
for $\protect\epsilon _{xy}=2i$ and $\protect\epsilon _{xy}=3.5i$,
respectively. The red surfaces represent chiral surface waves and the green
one is the vacuum band.}
\label{fig3}
\end{figure}

Because of the same topology, the edge states for $k_{z}>0$ and $k_{z}<0$
propagate along the same direction (Figs.~\ref{fig2}(b),\ref{fig3}(b)). We
see for a given $k_{x}$ and $\omega $ at the edge, there could be two
surface states with opposite $k_{z}$. In Fig.~\ref{fig2}(c), we show these
two edge modes along $k_{z}$ for a fixed $k_{x}$, which start from the lower
band and gradually approach the upper band. As a comparison, there may be
only one edge mode along $k_{z}$ for a fixed large $k_{x}$ with the chiral
effect \cite{Supp}. Such double edge modes originate from topologically
trivial 2D bands in the $k_{y}$-$k_{z}$ plane for a fixed $k_{x}$, which
gives zero or even numbers of edge modes with opposite chirality.

We remark that when both gyromagnetic and chiral effects are considered,
their competition would drive a transition between charge-2 and charge-0
TDPs. An example is shown explicitly in SM \cite{Supp}.

{\color{blue}\emph{Bulk-edge correspondence with vacuum bands. }}Unlike
electronic materials, vacuum is not an insulator for photons and there exist
photonic bands for vacuum (although topologically trivial), \textit{i.e.},
the free space continuum. Because of its direct contact with the edges of
HMMs, vacuum effects should be taken into account for edge states and
bulk-edge correspondence. Here we illustrate the vacuum effects using
gyromagnetic effects. For a small $\epsilon _{xy}$ term, the vacuum band is
higher than both bulk bands of the HMM (the vacuum band was not shown in
Fig. \ref{fig2}(b,c) for this reason). With increasing $\left\vert \epsilon
_{xy}\right\vert $, the band gap between the two nonzero bulk bands
increases and the upper band would surpass the vacuum bands at a certain
value of $\left\vert \epsilon _{xy}\right\vert $, after which the edge mode
connects to the vacuum band, instead of the upper band, as shown in Fig.~\ref%
{fig3}(a). This is because photons cannot localize at the boundary of the
HMM after they diffuse into vacuum. Since the vacuum band is topologically
trivial, the physical properties of the surface waves like chirality are
preserved.

In Fig.~\ref{fig3}(b,c), we plot the 3D band structures with edge modes for
both weak ($\epsilon _{xy}=2i$) and strong ($\epsilon _{xy}=3.5i$)
gyromagnetic effects in the $k_{z}$-$k_{x}$ plane, which show similar
features as Figs. \ref{fig2}(b) and~\ref{fig3}(a). Note that the vacuum band
crosses $(\mathbf{k},\omega )=0$ and does not intersect with the non-zero
upper HMM band away from the TDP because both bands increase linearly with
respect to $|k|$. The surface states, starting from the TDP, always fill the
common gap between the lower HMM band and either the upper HMM or vacuum
bands, depending on which has the lower energy. For a given $\epsilon _{xy}$%
, only one band (upper HMM band (Fig.~\ref{fig3}(b)) or vacuum band (Fig.~%
\ref{fig3}(c)) for weak and strong gyromagnetic effects, respectively) is
connected by the surface states.

\begin{figure}[h]
\centering
\includegraphics[width=0.48\textwidth]{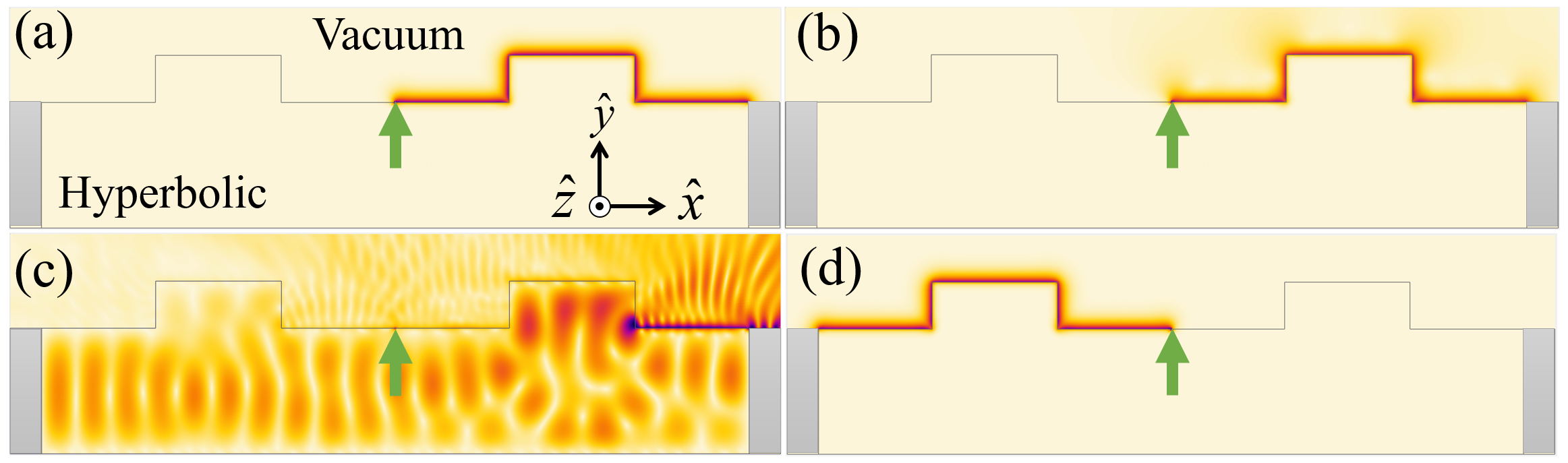}
\caption{COMSOL simulation results for our model, where a HMM is placed in
vacuum with two absorption materials on two sides (gray-coded areas). The
color represents the distribution of total electric field strength. The
green arrow indicates the position of a line source, which is a plane wave
along vertical direction, with input energy $\protect\omega _{I}$. The field
propagates along the $z$-direction. We choose the same parameters as those
in Fig.~\protect\ref{fig3}(a) and tune the input source to (a) $\protect%
\omega _{I}=0.95$, (b) $\protect\omega _{I}=1$ and (c) $\protect\omega %
_{I}=1.45$. (d) Same as panel (a) except that the sign of the applied
gyromagnetic term is opposite such that the chirality of the edge state is
reversed.}
\label{fig4}
\end{figure}

{\color{blue}\emph{Numerical simulations. }}The above topological band
properties and corresponding edge states in continuous HMMs can be further
confirmed through COMSOL Mutiphysics. Here we choose three different values
of line source frequency $\omega _{I}=0.9$, $1$ and $1.45$, which correspond
to band energies below the vacuum band, overlapping with the vacuum band,
and overlapping with both the vacuum and bulk bands, respectively (Fig.~\ref%
{fig3}(a)). The simulation results are shown in Fig.~\ref{fig4}. In panel
(a), when $\omega _{I}$ just lies below the vacuum band, the surface wave
moves along the positive direction and is robust to any scattering process.
When we increase $\omega _{I}$ a little bit so that it overlaps with the
vacuum band, the surface wave is scattered into vacuum at defective points
and source (panel (b)). If $\omega _{I}$ overlaps with both vacuum and bulk
bands, as well as the gapless surface state, the electromagnetic waves
diffuses into the entire space while the right side has a stronger field
intensity (panel (c)). Finally, since the chirality of edge states is
determined by sign($\Im (\epsilon _{xy})$), the surface wave indeed travels
along the opposite direction when the gyromagnetic term is changed to an
opposite sign in Fig.~\ref{fig4}(d).

Here, we mainly concern the simulations with gyromagnetic terms while the
chirality cases are studied in Supplementary Materials \cite{Supp}.

{\color{blue}\emph{Discussions and conclusion.}} We have considered a HMM
with hyperbolicity on the permittivity tensor, which, however, is not
necessary for the existence of chiral surface wave. For instance, a HMM with
$\epsilon =I$ and $(\mu _{x}>0,\mu _{y}>0,\mu _{z}<0)$ may also exhibit
chiral surface waves under proper time-reversal (or inversion) symmetry
breaking. Besides $\epsilon _{xy}$, the gyromagnetic effects can also be
generated by non-diagonal terms in $\mu $. Indeed, a purely imaginary $\mu
_{xy}$ induces chiral surface waves in a similar way, which, however,
becomes topologically trivial (gapless) upon passing the critical point $\Im
(\mu _{xy})=\pm \sqrt{\mu _{x}\mu _{y}}$ \cite{Supp}.

For experimental considerations, the chiral effects exist in a range of
natural materials \cite{ViitanenAJ1994} while the advances of metamaterials
allow us to synthesize strong chiral media \cite{OhSS2015}. To achieve
gyromagnetic effects, magnetic materials can be mixed during fabrication and
one commonly used material is Yttrium-Iron-Garnet \cite{WangZ2009}.

The topological band theory described here can be applied to various
parameter regions and many interesting effects, such as gain and loss \cite%
{Feng2017}, disorder, bianisotropy terms with more general $\gamma $ tensor,
remain to be explored. The hyperbolic band dispersion of the topological
HMMs opens a new avenue for studying negative refraction with topological
edge states as well as topological lasing. In particular, the topological
edge states in HMMs may be used to design a \textit{topological-semimetal
laser}. By tuning the structure of HMM
and gyromagnetic/chiral field, the topological edge mode can be promoted to
the lasing mode, rendering a highly efficient single-mode laser, which is
robust to local disorders and defects.
Note that although the bulk spectrum
of HMMs could be complex, the topological edge spectrum is purely real. Thus it
does not suffer from the inherent loss, which is the primary roadblock to the
insertion of bulk HMMs into practical technologies. Because of the important
and unique properties of HMMs like broad-band spontaneous emission
enhancement (thus, the lasing threshold would be very small) and the ability to support propagations
of large-momentum waves \cite{PoddubnyA2013}, the topological-semimetal
laser may outperform recently emerged topological insulator laser using
photonic crystals \cite{Harari2018,Bandres2018}.

In conclusion, we developed a topological band description for the
non-Hermitian continuous HMMs and found two types of non-Hermitian photonic
triply-degenerate points (classified by their topological charges) with
different surface states. Our work should provide physical understanding of
topological phases in HMMs and may inspire further theoretical and
experimental investigations on the fundamental properties as well as
practical applications of topological photonic continua.

\begin{acknowledgments}
\textbf{Acknowledgements}: We thank W. Gao for helpful discussions about
COMSOL simulations. This work was supported by Air Force Office of
Scientific Research (FA9550-16-1-0387), National Science Foundation
(PHY-1505496), Army Research Office (W911NF-17-1-0128) and UTD seed grant.
Z. Li and Q. Gu acknowledge funding from UT Dallas faculty start-up funding.
\end{acknowledgments}

\newpage \clearpage
\onecolumngrid
\appendix

\section{Supplementary Material for ``Topological bands and triply-degenerate
points in non-Hermitian continuous hyperbolic metamaterials''}
\setcounter{table}{0} \renewcommand{\thetable}{S\arabic{table}} %
\setcounter{figure}{0} \renewcommand{\thefigure}{S\arabic{figure}}
\def\theequation{S\arabic{equation}}

\subsection{Imaginary bands and non-Hermitian photonic TDP}

Here, we provide additional information about the imaginary band structure
and the emergence of various nontrivial photonic band touchings including
quadratic band touching, Dirac points and the non-Hermitian photonic TDP.

\begin{figure}[h]
\centering
\includegraphics[width=0.75\textwidth]{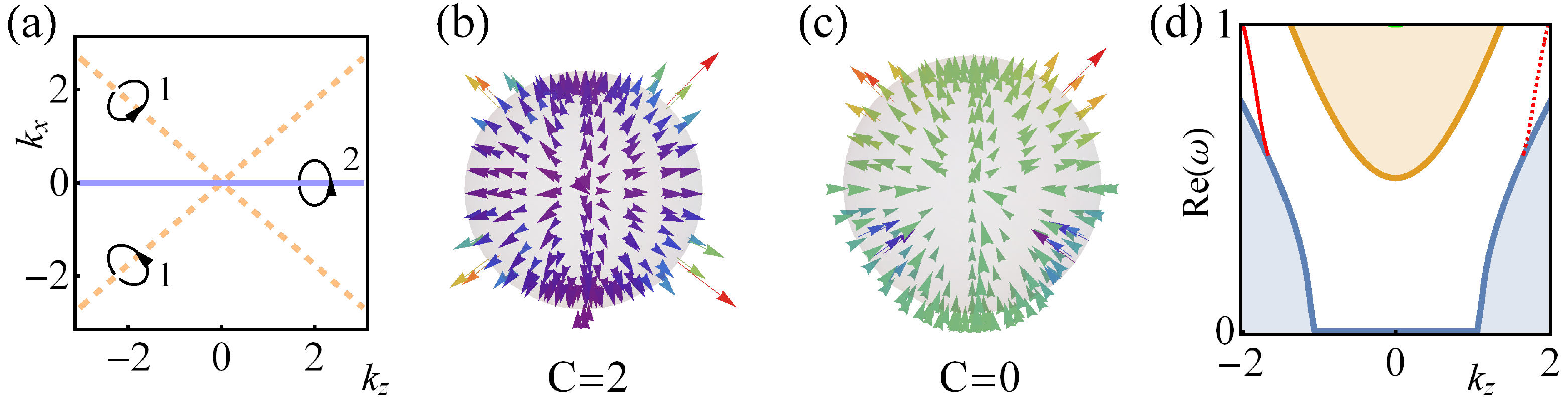}
\caption{(a) The degenerate line in Fig.~1(a) has a winding number 2 (solid
light-blue line). The line splits into two (dashed light orange lines) in
the $k_{y}=0$ plane for $(\protect\epsilon _{x},\protect\epsilon _{y},%
\protect\epsilon _{z})=(1,4,-3)$, each of which has a winding number 1.
(b,c) The distributions of Berry curvature on closed unit spheres containing
charge-2 and charge-0 TDPs. We choose $(\protect\epsilon _{x},\protect%
\epsilon _{y},\protect\epsilon _{z})=(4,4,-3)$. Two types of TDPs are
induced by $\protect\gamma =\text{diag}(1,0,0)$ and $\protect\epsilon _{xy}=-%
\protect\epsilon _{yx}=2i$, respectively. (d) Similar as Fig.~2(a), but the
edge states are calculated along the $k_{z}$ direction with $k_{x}=1$
(dashed red curve) or $k_{x}=-1$ (solid red curve).}
\label{figS2}
\end{figure}

As plotted in Fig.~1(a) in main text, the degenerate line has a winding
number 2. In fact, when $\epsilon _{x}>\epsilon _{y}$ ($\epsilon
_{x}<\epsilon _{y}$), the single line splits into two degenerate lines, each
carrying a winding $W=1$ in $k_{x}=0$ ($k_{y}=0$) plane. These has been
illustrated in the projected 2D plane in Fig.~\ref{figS2}(a), where the
corresponding winding number $W$ has been labelled. Regardless of the number
and the winding of the degenerate lines, they can be gapped out by either
chiral or gyromagnetic effects, rendering two types of TDPs, whose
topological charge distributions (Berry curvatures) are plotted in Fig.~\ref%
{figS2}(b) and (c) respectively.

For the charge-2 TDP (Fig.~\ref{figS2}(b)), four regions with outgoing Berry
flux contribute significantly to the total Chern number while the Berry
curvature in other regions is close to 0. For the charge-0 TDP (Fig.~\ref%
{figS2}(c)), the Berry curvature is still significant in the four regions,
but the outgoing Berry flux on the top half of the sphere cancels with the
incoming flux on the bottom half, yielding zero net Berry flux, thus
vanishing topological charge. Different from the charge-0 TDP, the 2D band
topology for the charge-2 TDP is also non-trivial in the $k_{y}$-$k_{z}$
plane (with fixed $k_{x}\neq 0$). For a large $|k_{x}|$, we may only find
one chiral surface wave at $k_{z}=1$ ($k_{z}=-1$) represented by the dashed
(solid) red curve in Fig.~\ref{figS2}(d).

\begin{figure}[h]
\centering
\includegraphics[width=0.65\textwidth]{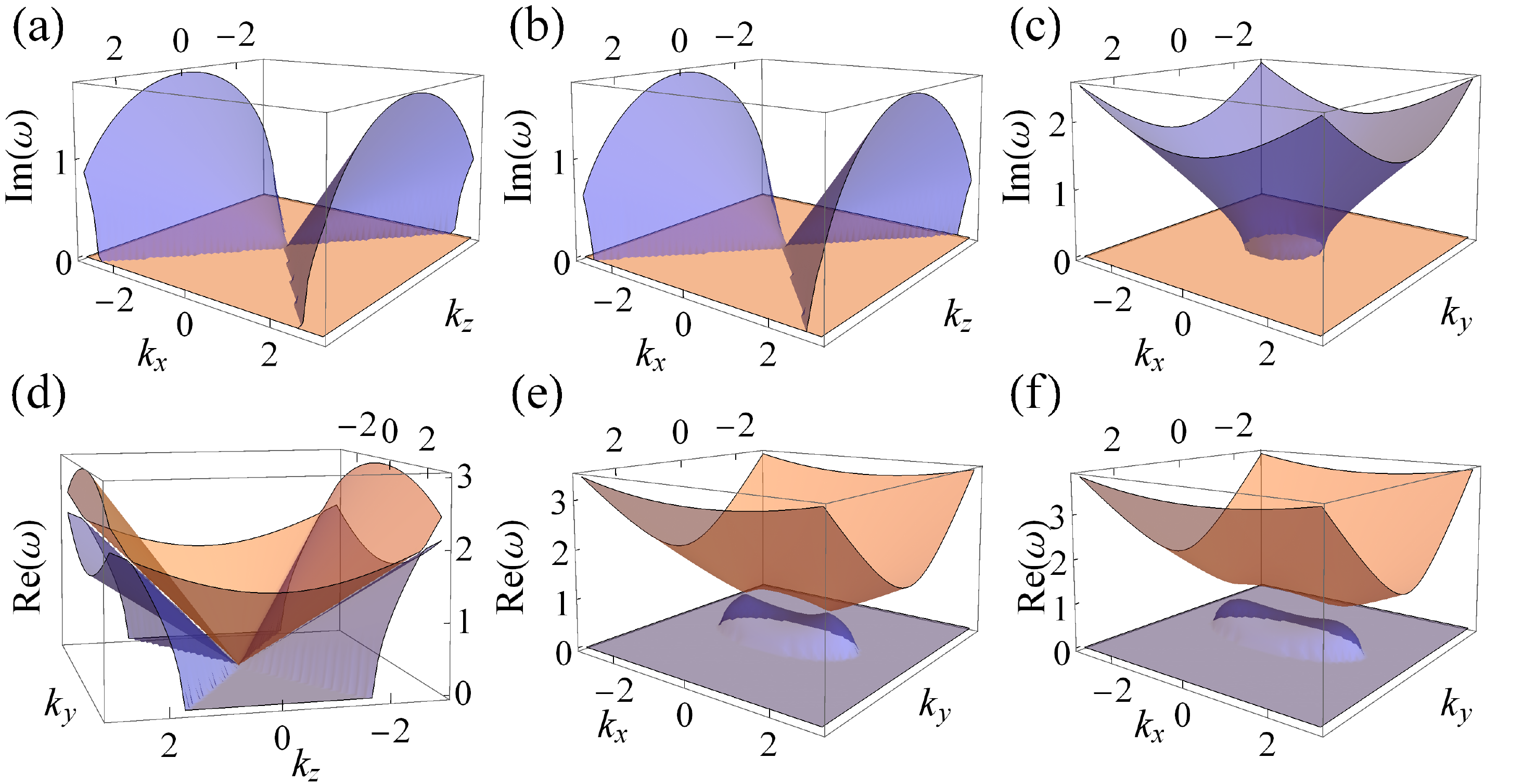}
\caption{(a,b,c) The imaginary bands for Figs.~1(a,b,c) in the main text.
(d,e) Two winding 1 degenerate lines and corresponding Dirac points at $%
k_{z}=1$ for the HMM with $(\protect\epsilon _{x},\protect\epsilon _{y},%
\protect\epsilon _{z})=(1,4,-3)$. (f) The degenerate lines and Dirac points
in (d,e) can be gapped by proper fields and two bands become topologically
non-trivial.}
\label{figS1}
\end{figure}

In Fig.~1(a,b) in the main text, we show that the degenerate line can be
lifted by the chirality term. The chiral effect only changes the real parts
of the bands and leaves the imaginary parts unchanged, as shown in Fig.~\ref%
{figS1}(a,b). The upper band (orange) is purely real (i.e., $\Im (\omega )=0$%
) and the lower band changes from purely real to purely imaginary at two
lines crossing the TDP. Fig.~\ref{figS1}(c) shows the imaginary spectrum for
Fig.~1 (c), where the upper band is purely real in the entire $k_{x}$-$k_{y}$
plane. For any finite $k_{z}$, the lower band changes from purely real to
purely imaginary across an exceptional ring. The exceptional ring forms a
exceptional degenerate cone in the 3D momentum space (see Fig.~1(d) in main
text and also Fig.~\ref{figS3}), which does not prevent defining the band
Chern number of the lower band as we will discuss later.

When the winding 2 degenerate line splits into two degenerate lines with
winding number 1 (see Fig.~\ref{figS2}(a)), the quadratic touching point
splits into two charge-1 Dirac points as well, which are depicted in Fig.~%
\ref{figS1}(e). Similarly, the two Dirac points can be gapped  with either
gyromagnetic or chirality effects (Fig.~\ref{figS1}(f)), leading to a
nontrivial phase in $k_z\neq0$ plane.

\begin{figure}[h]
\centering
\includegraphics[width=0.85\textwidth]{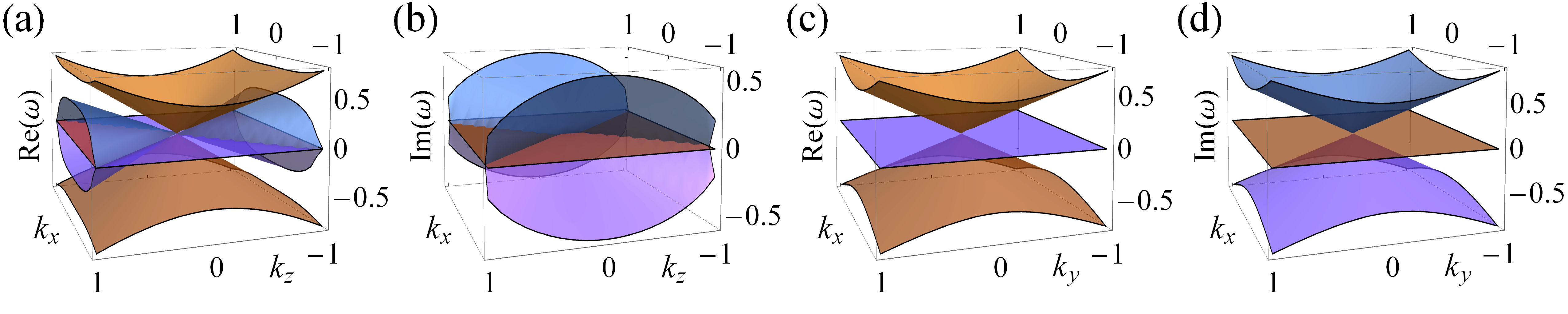}
\caption{(a) and (b) Real and imaginary spectra in the $k_{x}$-$k_{z}$ plane
with $k_{y}=0$. (c) and (d) Real and imaginary spectra in the $k_{x}$-$k_{y}$
plane with $k_{z}=0$. The parameters are $\protect\epsilon =(4,4,-3)$, $%
\protect\mu =1$ and $\protect\gamma =(1,0,0)$.}
\label{FigSFB}
\end{figure}

So far we only plot the two relevant bands for simplicity. To better
visualize the non-Hermitian TDP, we plot the full six bands for a
chirality-induced charge-2 photonic TDP in Fig.~\ref{FigSFB}. Clearly, the
full bands, both real and imaginary parts, are symmetric to $\omega =0$ as
expected.

We first consider the $k_{y}=0$ plane and the real/imaginary spectra are
plotted in Figs.~\ref{FigSFB}(a,b). The top/bottom band is purely real and
shows linear dispersions along arbitrary directions in the plane. The middle
dispersive bands form a cone structure along $k_{z}$/$k_{x}$ in the
real/imaginary part, which is distinct from the Hermitian TDPs in either
electronic \cite{TDP2} or atomic \cite{HuH2018} systems. Similar bands are
observed at the $k_{x}=0$ plane because the type-I HMM shows dielectric
properties in both $x$ and $y$ directions.

For the $k_{z}=0$ plane, the photonic TDP also mixes both real and imaginary
bands, but does not show the cone structure. As shown in Figs.~\ref{FigSFB}%
(c,d), the middle flat band is two-fold degenerate. For the top and bottom
bands, each has a purely real and purely imaginary branches. All of them are
linear along any direction in the $k_{x}$-$k_{y}$ plane.

\subsection{Band gap and quantization of Chern number}

As discussed in the previous section, there is a TDP locating at the origin
with proper gyromagnetic/chiral fields. When $k_{z}$ deviates from 0, the
degeneracy is lifted, which is essential for the non-trivial topology.
However, we have not proven that the TDP is the only degeneracy along $k_{z}$%
.

We prove this by analytically finding the energy gap. We first consider the
gyromagnetic case with non-zero $\epsilon _{xy}=ia_{\epsilon }$ and $%
\epsilon _{yx}=-ia_{\epsilon }$. Without loss of generality, we take $%
\epsilon _{x}=\epsilon _{y}>0$ and $\epsilon _{z}<0$ and find the gap size
is proportional to $|k_{z}/\sqrt{a_{\epsilon }+\epsilon _{x}}|$. Similarly,
we can find the band gap $\left\vert k_{z}\frac{\sqrt{\epsilon
_{x}^{3}-a_{\gamma }^{2}\epsilon _{x}+\sqrt{\epsilon _{x}^{3}(a_{\gamma
}^{3}-\epsilon _{x}a_{\gamma })^{2}}}}{(a_{\gamma }^{2}-\epsilon
_{x})\epsilon _{x}}\right\vert $ for non-zero chiral effects along the $x$
direction $\gamma _{x}=a_{\gamma }>0$. In both cases, the gaps are
proportional to $|k_{z}|$ (linear dependence), therefore would not close at
any finite $k_{z}$.

With a well-defined band gap, we can compute the topological invariant of
each band. For a continuous system, the band Chern number can be quantized
when the integration range goes to infinity in the momentum space. Here we
first provide the numerical results to illustrate the quantization of Chern
number. We consider the same case as Fig.~2(a) in the main text and compute
the Chern number in different regimes as shown in Tab.~\ref{tab1}. The
integral range is set as $-k_{max}\leq k_{x},k_{y}\leq k_{max}$ and we find
that the Chern number indeed converges to a quantized number at sufficiently
large $k_{max}$. By inspecting the distribution of Berry curvature, we find
that the main contribution comes near the band gap opening point and the
Berry curvature decreases exponentially at large momentum.

\begin{table}[tbp]
\begin{tabular}{l|lllllll}
$k_{max}$ & $3\times10^0$ & $3\times10^1$ & $3\times10^2$ & $3\times10^3$ & $%
3\times10^4$ & $3\times10^5$ & $3\times10^6$ \\ \hline
$C$ & \textbf{0.963996} & \textbf{0.997366} & \textbf{0.999738} & \textbf{%
0.999974} & \textbf{0.999997} & \textbf{1.000000} & \textbf{1.000000}%
\end{tabular}%
\caption{Convergence of Chern number of upper band when the integration
range goes to large momentum. The system parameters are $(\protect\epsilon %
_{x},\protect\epsilon _{y},\protect\epsilon _{z})=(4,4,-3)$, $\protect\mu =1$
and $\protect\gamma =\text{diag}(1,0,0)$. The Chern number is computed on a
plane with constant $k_{z}>0$.}
\label{tab1}
\end{table}

While the above numerics only suggest the convergence of Chern number, we
can prove that Chern number is strictly quantized in certain cases. In order
to show this, we need to prove that the Berry curvature vanishes at
infinity, specifically, when $k_{z}=k_{z,0}$ is a finite non-zero constant
and $k_{x},k_{y}\rightarrow \infty $. It is mathematically challenging to
solve the eigenmodes directly and also hard to get any insight due to the
complex expressions. In order to circumvent this difficulty, we unitize the
linear $\mathbf{k}$ dependence to convert the eigenvalue problem at infinite
momenta to finite momenta. To show how this procedure can be performed, we
consider only right eigenvectors (left eigenvectors are similar) and rewrite
the eigenvalue problem in cylindrical coordinate as
\begin{equation}
H(k_{r}\cos \theta ,k_{r}\sin \theta ,k_{z,0})|\Psi \rangle =\omega |\Psi
\rangle ,
\end{equation}%
where we do not write the $\mathbf{k}$-dependence of $|\Psi \rangle $ or $%
\omega $ explicitly. The cylindrical coordinate is defined by $k_{r}=\sqrt{%
k_{x}^{2}+k_{y}^{2}}$ and $\tan \theta =k_{y}/k_{x}$. Due to the linearity
of ${p}(\mathbf{k})$, we have ${p}(k_{r}\cos \theta ,k_{r}\sin \theta
,k_{z,0})=k_{r}{p}(\cos \theta ,\sin \theta ,k_{z,0}/k_{r})$. The above
equation can be rewritten as
\begin{equation}
H(\cos \theta ,\sin \theta ,k_{z,0}/k_{r})|\Psi \rangle =(\omega
/k_{r})|\Psi \rangle ,
\end{equation}%
which becomes
\begin{equation}
H(\cos \theta ,\sin \theta ,0)|\Psi \rangle =\omega _{0}|\Psi \rangle
\label{EigenEqu}
\end{equation}%
when $k_{r}\rightarrow \infty $. Here $\omega _{0}=\lim_{k_{r}\rightarrow
\infty }\omega (k_{r})/k_{r}$ is a constant since $\omega (k_{r})$ must grow
linearly when $k_{r}$ is sufficiently large. Since the problem for infinite
momentum is converted to one for finite momentum, every quantity can be
computed exactly.

As an illustration, we consider an example with the following parameters: $%
\epsilon _{x}=\epsilon _{y}=-\epsilon _{z}>0$, $\epsilon _{xy}=-\epsilon
_{yx}=i\chi $, $\mu =\mu _{0}$ and $\gamma =0$. We only consider the case $%
\chi ^{2}<\epsilon ^{2}$ so that the upper band $\omega =\sqrt{\frac{%
\epsilon _{x}}{\mu _{0}}}\sqrt{\frac{k_{x}^{2}+k_{y}^{2}}{\epsilon
_{x}^{2}-\chi ^{2}}}$ is purely real and fully gapped from other bands. Note
that, once we impose $k_{x}=\cos \theta $ and $k_{y}=\sin \theta $, the
eigenfrequency $\omega $ reduces to a constant $\omega _{0}=\sqrt{\frac{%
\epsilon _{x}}{\mu _{0}}}\frac{1}{\sqrt{\epsilon _{x}^{2}-\chi ^{2}}}$ as in
Eq.~\ref{EigenEqu}. The corresponding normalized right eigenvector is
\begin{equation}
|\Psi (\theta )\rangle =\frac{1}{\epsilon _{x}^{2}+\epsilon _{x}\mu
_{0}-\chi ^{2}}\left( 0,0,-\frac{k_{x}\epsilon _{x}}{\omega },-\frac{%
k_{x}k_{y}(\epsilon _{x}^{2}-\chi ^{2})}{k_{x}^{2}+k_{y}^{2}},\frac{%
k_{x}^{2}(\epsilon _{x}^{2}-\chi ^{2})}{k_{x}^{2}+k_{y}^{2}},0\right) ^{T},
\end{equation}%
which leads to vanishing Berry curvature
\begin{equation}
\mathscr{F}=\langle \partial _{k_{x}}\Psi (\theta )|\partial _{k_{y}}\Psi
(\theta )\rangle -\langle \partial _{k_{y}}\Psi (\theta )|\partial
_{k_{x}}\Psi (\theta )\rangle =0
\end{equation}%
at any given momentum. This is equivalent to that the Berry curvature
vanishes at infinity along any path on the $k_{x}$-$k_{y}$ plane. In this
sense, the $\mathcal{R}^{2}$ momentum space (at finite $k_{z}$) can be
compactified into a sphere $S^{2}$, on which the topological classification
is the same as that on a tours and the quantization of Chern number is
guaranteed. Similar results can be drawn when we have gyromagnetic or chiral
effects.

\subsection{Topological transitions between two types of TDP}

\begin{figure}[h]
\centering
\includegraphics[width=0.9\textwidth]{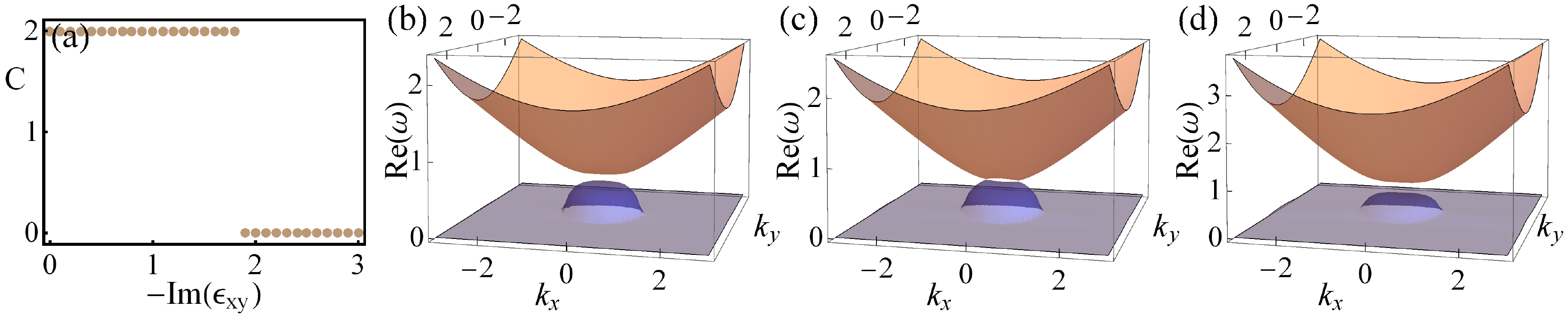}
\caption{(a) The charge of the TDP as a function of $\protect\epsilon_{xy}$
with fixed $\protect\gamma=\text{diag}(1,0,0)$. The TDP changes from charge
2 to charge 0 around $-\Im(\protect\epsilon_{xy})=1.85$. (b-d) The 2D band
structures in the $k_z=1$ plane with $-\Im(\protect\epsilon_{xy})=1$, $1.85$%
, $3$ respectively. Other parameters are $(\protect\epsilon_x,\protect%
\epsilon_y,\protect\epsilon_z)= (4,4,-3)$, $\protect\mu=I$.}
\label{figS2.5}
\end{figure}

When the chiral and gyromagnetic effects coexist, the competition between
them could drive a transition of the TDP between charge-2 and charge-0. In
Fig.~\ref{figS2.5}(a), we plot how the TDP charge changes as a function of
the gyromagnetic term $\epsilon _{xy}$ with fixed chiral term $\gamma =\text{%
diag}(1,0,0)$. The transition is closely related to the 2D band topology in
the $k_{x}$-$k_{y}$ plane with a fixed $k_{z}$. As we increase $|\epsilon
_{xy}|$ (consider $\Im \epsilon _{xy}<0$), the corresponding 2D Chern number
remains unchanged for $k_{z}<0$, but changes sign for $k_{z}>0$ ($k_{z}<0$).
At the transition point, the 2D band gap closes as shown in Figs.~\ref%
{figS2.5}(b-d), leading to the appearance of degenerate lines in the $%
k_{z}>0 $ region.

\subsection{Chiral surface wave by chirality effects}

\begin{figure}[h]
\centering
\includegraphics[width=0.75\textwidth]{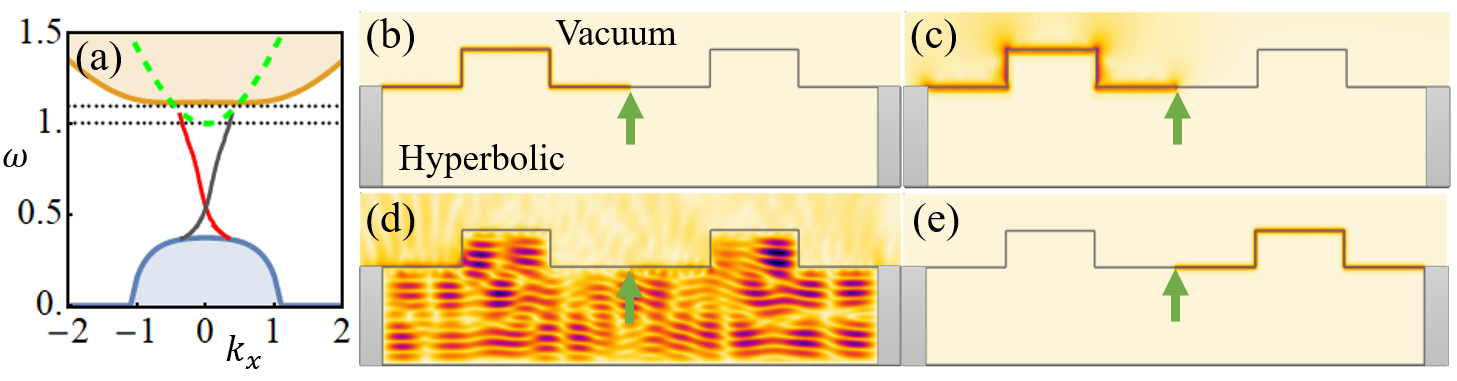}
\caption{(a) Similar to Fig.~2(a) but with chirality $\protect\gamma =\text{%
diag}(1.6,0,0)$. (b)-(d) COMSOL simulation results for panel (a) with $%
\protect\omega _{I}=0.6$, $1$, $1.1$ respectively. $k_{z}=1$. (e) Similar as
panel (b) but for $k_{z}=-1$.}
\label{figS6}
\end{figure}

With increasing chirality, the HMMs bands may also become higher than vacuum
bands. Here we take $\gamma =\text{diag}(1.6,0,0)$ for simulations and the
corresponding 2D band structure is plotted in Fig.~\ref{figS6}(a).

We consider three different frequencies in COMSOL simulations, similar as
those for gyromagnetic effects in the main text. $\omega _{I}=0.6$ lies
within the common gap and we observe the scattering-free chiral surface wave
as shown in Fig.~\ref{figS6}(b), whose direction is consistent with the
prediction of band theory. As we increase $\omega _{I}$ to 1, the frequency
cuts the vacuum band and the surface wave diffuses into vacuum at defects
(Fig.~\ref{figS6}(c)). $\omega _{I}=1.1$ overlaps with both bulk and vacuum
bands and there are no chiral surface waves (Fig.~\ref{figS6}(d)). Since the
chiral effects break inversion symmetry, the surface wave has opposite
chirality at $k_{z}=-1$, which is also confirmed by the simulation results
in Fig.~\ref{figS6}(e).

\subsection{Exceptional degeneracy of the lower band}

\begin{figure}[h]
\centering
\includegraphics[width=0.5\textwidth]{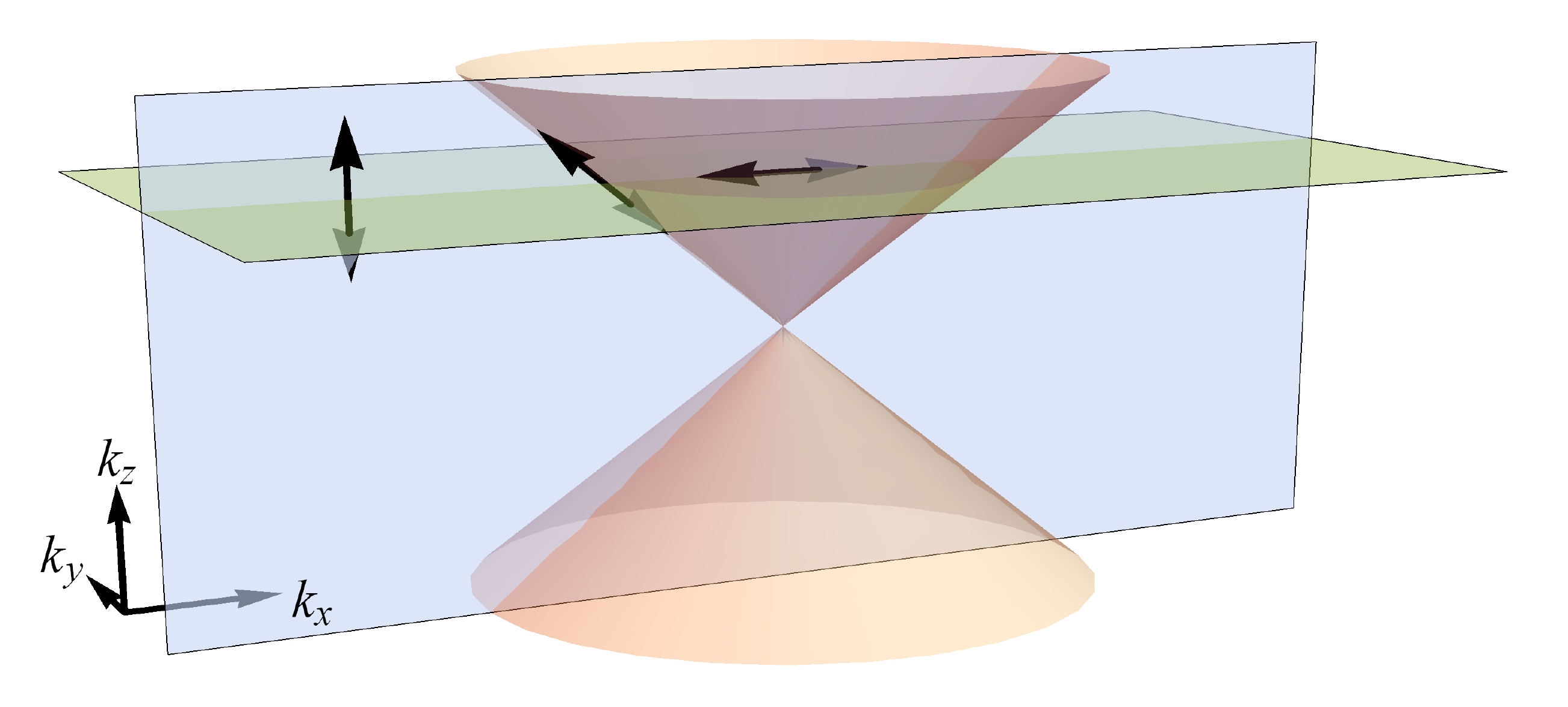}
\caption{Illustration of the exceptional cones (see also Fig.~1 (d)) in the momentum space between the zero flat
band and lower non-zero band. The arrows show the directions of the
polarizations at different $\mathbf{k}$, which become parallel with $\mathbf{%
k}$ on the cone.}
\label{figS3}
\end{figure}

As we discussed in the main text, away from $\mathbf{k}=0$, the upper band
is fully separated from both the lower band and the zero-energy flat band,
therefore its topology is well defined. However, the lower band may have
zero energy solutions, which are degenerate with the zero-energy flat band
in the momentum space. As shown in Fig.~1(c) and Fig.~\ref{figS1}(e,f), the
lower band (blue) would change from purely real to purely imaginary on an
exceptional ring where it is degenerate with the zero energy flat band. Such
an exceptional ring exists at any finite $k_{z}$ plane and thus forms an
exceptional (degenerate) cone in the entire momentum space (see Fig.~1 (d)
as well as Fig.~\ref{figS3}).

To understand such degeneracy, we consider the HMMs without chiral or
gyromagnetic effects. Without loss of generality, we consider the line in
the $k_{y}=0$ plane with a fixed $k_{z}$. As $k_{x}$ changes from $0$ to $%
-\infty $, the polarization (linear) of the lower-band states rotates in an
opposite direction with respect to the rotation of $\mathbf{k}$ (see Fig.~%
\ref{figS3}) due to the metal properties along the $z$ direction, which
becomes parallel with $\mathbf{k}$ at some critical $k_{x}$, leading to the
zero energy solution through $\nabla \times \bm E=0$. As $k_{x}$ changes
from $0$ to $-\infty $, the lower-band eigenstates change continuously, even
across the degenerate points. Therefore these degeneracies do not affect the quantization of the Chern number.

In the presence of chiral or gyromagnetic effects, the changes of the
polarization are similar although it is no longer linear. As $k_{x}$ changes
from $0$ to $-\infty $, the polarization changes from left-handed
(right-handed) elliptical polarization to right-handed (left-handed)
elliptical polarization. At the degenerate point, the polarization is linear
and parallel with $\mathbf{k}$. Moreover, we also notice that Berry
curvatures of the lower band mainly resides on the real-valued energy region.

\subsection{EFS theory: connection to band theory and its incompleteness}

In this subsection, we discuss how the EFS theory \cite%
{LiuC2017,GaoW2015,ChernRL2008} relates to our topological band theory and
why EFS is not suitable for studying topological properties, although it may
happen to provide some correct predictions in certain parameter regimes.
Finally, we will show a counterexample where the EFS theory gives misleading
answers while the prediction of our band theory is correct.

First of all, in both frameworks, the edge states are the same physical
states, which are solved from matching boundary condition of Maxwell's
equations along the open-boundary direction. Here we choose it to be the $y$
direction. The electric fields are solved from the master equation
\begin{equation}
\bm{K}\times \mu ^{-1}\bm{K}\times \bm{E}+i\omega (\bm{K}\times \mu
^{-1}\gamma \bm{E}+\gamma \mu ^{-1}\bm{K}\times \bm{E})+\omega ^{2}(\epsilon
-\gamma \mu ^{-1}\gamma )\bm{E}=0,
\end{equation}%
which can be easily derived from Maxwell's equations and the constitutive
relations listed in the main text.

The topological invariant used in the EFS theory is also Chern number. In
general, there are three Chern numbers, corresponding to three EFSs (the
middle one may disappear in certain cases, leaving only two EFSs). The
middle EFS (Chern number $C_{1}$) is usually a simple closed 2-manifold
while the other two EFSs ($C_{2}$) stretch to infinity, which are defined by
the upper band and lower band in our band theory, respectively. When the
middle one contains the origin $\mathbf{k}=0$, it simply gives the charge of
the TDP. As a result, we expect to see $|C_{1}|=2$ and $C_{1}=0$ for
chirality and gyromagnetic effects, respectively. In addition, $|C_{2}|=1$
since it should be identical to the 2D Chern number calculated by the real
part of lower band in the $k_{x}$-$k_{y}$ plane with a fixed $k_{z}$. An
example of EFSs with Chern number and edge states are plotted in Fig.~\ref%
{figS4}(a).

In this sense, EFSs indeed capture certain information of band topology.
However, they are insufficient to fully characterize topological properties
of HMMs due to the lack of well-defined bulk-edge correspondence and their
inability of describing topological phase transitions. In the band theory,
we require a gapped band structure to define topological invariant on each
band and the difference between band Chern numbers will give the number of
edge modes (the Chern number must sum over all bands below Fermi surface for
a fermionic system). Nevertheless, there is no direct connection between the
number of chiral edge modes and Chern number of Fermi surfaces. Topological
phase transitions are usually signaled by gap closing, while the EFS theory
may not capture them due to the non-Hermitian nature of HMMs where the gap
closing may happen in the complex plane.

\begin{figure}[h]
\centering
\includegraphics[width=0.8\textwidth]{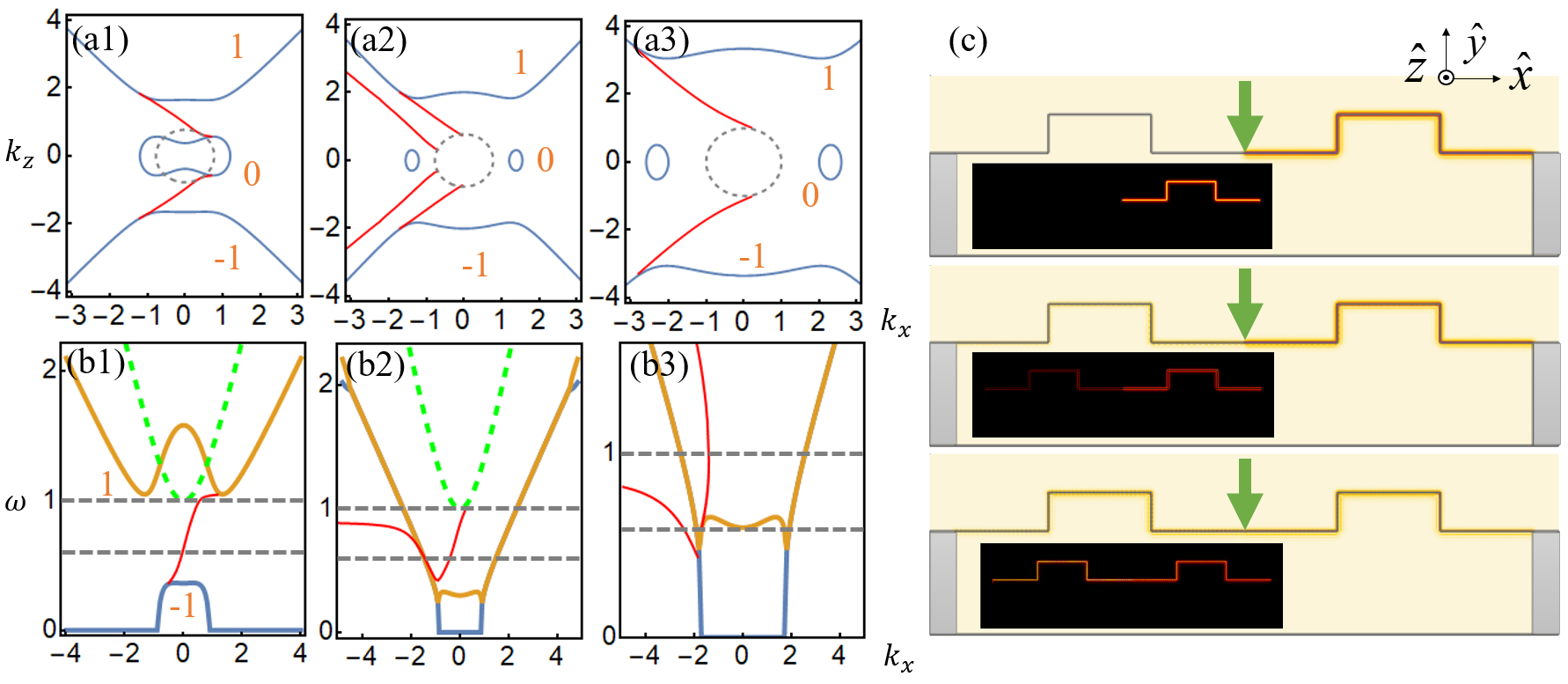}
\caption{Comparison between EFS and band theory. We choose HMMs with $(%
\protect\epsilon_x,\protect\epsilon_y,\protect\epsilon_z)=(4,4,-3)$ and the
same open-boundary condition in main text. The gyromagnetic effect is
incorporated through $\protect\mu _{xy}=-\protect\mu _{yx}$. (a) The
predictions given by EFS with (a1) $\protect\mu _{xy}=0.9i$, $\protect\omega %
=0.6$ , (a2) $\protect\mu _{xy}=1.8i $, $\protect\omega =0.6$ and (a3) $%
\protect\mu _{xy}=1.8i$, $\protect\omega =1$. The solid blue curves, solid
red curves and dashed gray curves are EFS of HMMs, edge states and EFS of
vacuum. The Chern number of each EFS is labelled. In all three cases, the
Chern numbers of EFSs are the same. (b) 2D band structure with (b1) $\protect%
\mu _{xy}=0.9i$, $k_{z}=1$, (b2) $\protect\mu _{xy}=1.8i$, $k_{z}=1$, (b3) $%
\protect\mu _{xy}=1.8i$, $k_{z}=2$. Only in (b1), the 2D bands are gapped
with non-trivial band Chern number. The dashed lines in (b2,b3) are $\protect%
\omega =0.6$, $1$ used for obtaining EFS. (c) COMSOL simulation results at $%
\protect\omega _{I}=0.6$. From top to bottom, we set $\protect\mu _{xy}=0.9i$%
, $\protect\mu _{xy}=1.1i$ and $\protect\mu _{xy}=1.8i$, respectively. The
surface wave is chiral only at $\protect\mu _{xy}=0.9i$ while the other two
are trivial edge modes. The insets give high-contrast images since the
non-chiral surface wave has weak electric fields. The simulation results
conflict with the EFS predictions, but are in agreement with our band-theory
descriptions. }
\label{figS4}
\end{figure}

Specifically, we find that when a large gyromagnetic effect is introduced
via permeability, the band topology becomes trivial with a vanishing band
gap. Consequently, there should be no chiral edge state according to our
band theory, which is also confirmed by COMSOL simulations. However, the EFS
theory still predicts the existence of topological edge states in such
parameter region.

We start from a HMM with the permittivity tenor $(\epsilon_x,\epsilon_y,%
\epsilon_z)= (4,4,-3)$ and permeability tensor $\mu =I$. The degeneracy in
Fig.~1 (a,b) are lifted by the non-diagonal terms $\mu _{xy}=-\mu _{yx}$,
which are both purely imaginary. As we indicated in the discussion section
of the main text, there is a phase transition point $\mu _{xy}=i$, at which
the permeability tensor in not invertible. Therefore we take $\mu _{xy}=0.9i$
and $1.8i$ across the transition and show the EFSs in Fig.~\ref{figS4}(a).
In both cases, the EFSs Chern numbers (from top to bottom) are calculated to
be 1, 0 and $-1$. The corresponding edge state solutions are also plotted.
The existence of non-vanishing Chern number suggests that the system should
be topologically non-trivial for all $\mu _{xy}$.

For $\Im (\mu _{xy})<1$, the EFS theory actually gives the correct
prediction (Fig.~\ref{figS4}(a1)). However, for $\mu _{xy}=1.8i$, the EFS
theory does not give consistent results for different energies. Two examples
with $\omega =0.6$ and 1 are plotted in Fig.~\ref{figS4}(a2,a3). In Fig.~\ref%
{figS4}(a2), there exist two surface waves (one starts from the vacuum),
which do not agree with the EFS Chern number and spoil the bulk-edge
correspondence in the EFS theory. Furthermore, the results in Fig.~4(a3) for
$\omega =1$ may lead to the impression that it is a topologically
non-trivial phase because the surface wave seems to match with the EFS Chern
numbers. These types of results lead to incorrect conclusions drawn by
authors of \cite{ChernRL2008}.

We investigate the same parameter region using the topological band theory.
In Fig.~\ref{figS4}(b1), we show the 2D band structure at $\mu _{xy}=0.9i$
and $k_{z}=1$, which is indeed topologically non-trivial. Nevertheless, as
the gyromagnetic term passes the critical transition point, the band becomes
gapless and the band Chern number cannot be defined.
Consequently, the system must be trivial and there is no chiral surface wave
anymore.

A natural question is why there is only one edge state solution in Fig.~\ref%
{figS4}(a3), while two in Fig.~\ref{figS4}(a2)? We explore this question
through solving the 2D band structure and the corresponding surface waves
with a fixed $\mu _{xy}$ but different $k_{z}=1,2$. The results are plotted
in Figs.~\ref{figS4}(b2,b3). In Fig.~\ref{figS4}(b2), we see a surface wave
starts from the vacuum band, inverts its chirality at certain $k_{x}$, then
passes through the imaginary bands of the HMM, and finally approaches $%
\omega =1$ at infinity. For $k_{z}=2$ shown in panel \ref{figS4}(b3), the
surface wave is split by the real bands of the HMM and the end of vertical
one intersects with the dash line $\omega =0.6$. This reveals that two
solutions in Fig.~\ref{figS4}(a2) actually correspond to one surface wave,
which exhibits opposite chirality in different regions of the momentum
space. Therefore the unidirectionality of surface wave is spoiled for $\mu
_{xy}=1.8i$. The COMSOL simulations also support such a conclusion, as shown
in Fig.~\ref{figS4}(c). From top to bottom, we show the total electric field
strength at $\mu _{xy}=0.9i$, $1.1i$ and $1.8i$ respectively. Only the first
one exhibits chiral surface wave while the remaining two are trivial because
the edge states propagate along both directions like plasma surface wave.

Nevertheless, even the surface wave solutions are not taken into
consideration, the EFS theory already does not work because it gives
non-trivial topological invariants in a trivial region and does not capture
the topological phase transition.


\begin{thebibliography}{99}
\bibitem{note} Note that the term ``continuous" here is defined in the
context of long wave limit, where the period for the lattice structure to
generate hyperbolic metametrials is much shorter than the optical wavelength.
\end{thebibliography}
\end{document}